\documentclass[letterpaper, 10 pt, conference]{ieeeconf}  

\IEEEoverridecommandlockouts                              

\usepackage{graphics} 
\usepackage{float}  
\usepackage{epsfig} 
\usepackage{mathptmx} 
\usepackage{times} 
\usepackage{amsmath} 
\usepackage{amssymb}  
\usepackage{xcolor}

\usepackage{hyperref}
\hypersetup{
    colorlinks=true,
    linkcolor=blue,
    filecolor=magenta,      
    urlcolor=cyan,
}

\title{\LARGE \bf
QuickTumorNet: Fast Automatic Multi-Class Segmentation of Brain Tumors}
\author{Benjamin Maas$^{1}$, Erfan Zabeh$^{1}$, and Soroush Arabshahi$^{1}$

\thanks{$^{1}$ Department of Biomedical Engineering, Columbia University,
        New York, NY, USA.  {\tt\small bjm2188@columbia.edu}}}%
        
\begin{document}

\maketitle
\thispagestyle{empty}
\pagestyle{empty}

\begin{abstract}
 Non-invasive techniques such as magnetic resonance imaging (MRI) are widely employed in brain tumor diagnostics. However, manual segmentation of brain tumors from 3D MRI volumes is a time-consuming task that requires trained expert radiologists. Due to the subjectivity of manual segmentation, there is low inter-rater reliability which can result in diagnostic discrepancies. As the success of many brain tumor treatments depends on early intervention, early detection is paramount. In this context, a fully automated segmentation method for brain tumor segmentation is necessary as an efficient and reliable method for brain tumor detection and quantification. In this study, we propose an end-to-end approach for brain tumor segmentation, capitalizing on a modified version of QuickNAT, a brain tissue type segmentation deep convolutional neural network (CNN). Our method was evaluated on a data set of 233 patient's T1 weighted images containing three tumor type classes annotated (meningioma, glioma, and pituitary). Our model, QuickTumorNet, demonstrated fast, reliable, and accurate brain tumor segmentation that can be utilized to assist clinicians in diagnosis and treatment.

    
   

\end{abstract}

\section{INTRODUCTION}

Fast brain tumor detection and classification is extremely important in patient outcomes. Speed and accuracy are of utmost importance regarding brain tumor detection, quantification, and classification. Accurate classification or localization of brain tumor types and tumor volume quantification are primary measures that dictate treatment and patient care ultimately determining the success of treatment and the patient's life expectancy. In other words, early intervention and appropriate treatment in brain tumors are dependent on tumor detection, classification, and quantification which directly relate to patients’ overall prognoses~\cite{liang2020prognostic,buerki2018overview,mete2017overview}. In our study, we look at three main brain tumor classifications. Meningioma, glioma, and pituitary brain tumors. Meningiomas surface on and stem from any intracranial or spinal dural surface, like pia, arachnoid, or dura mater. While believed to be generally benign, they can cause dramatic symptoms leading to a drastic decrease in quality of life~\cite{buerki2018overview}. Gliomas tumors stem from the central nervous system's (CNS) and peripheral nervous system's (PNS) supportive cells, neuralgia. Patients with glioma tumors prognosis are generally poor, particularly for patients with tumors that are invasive and malignant in nature~\cite{liang2020prognostic}. Pituitary tumors materialize on the pituitary gland and depending on their size and location can result in different hormonal effects~\cite{mete2017overview}. Each tumor type classification has respective implications for treatments, therapies, and patient care~\cite{liang2020prognostic,buerki2018overview,mete2017overview}. 
   \begin{figure}[h]
      \centering
      \includegraphics[width=0.4\textwidth]{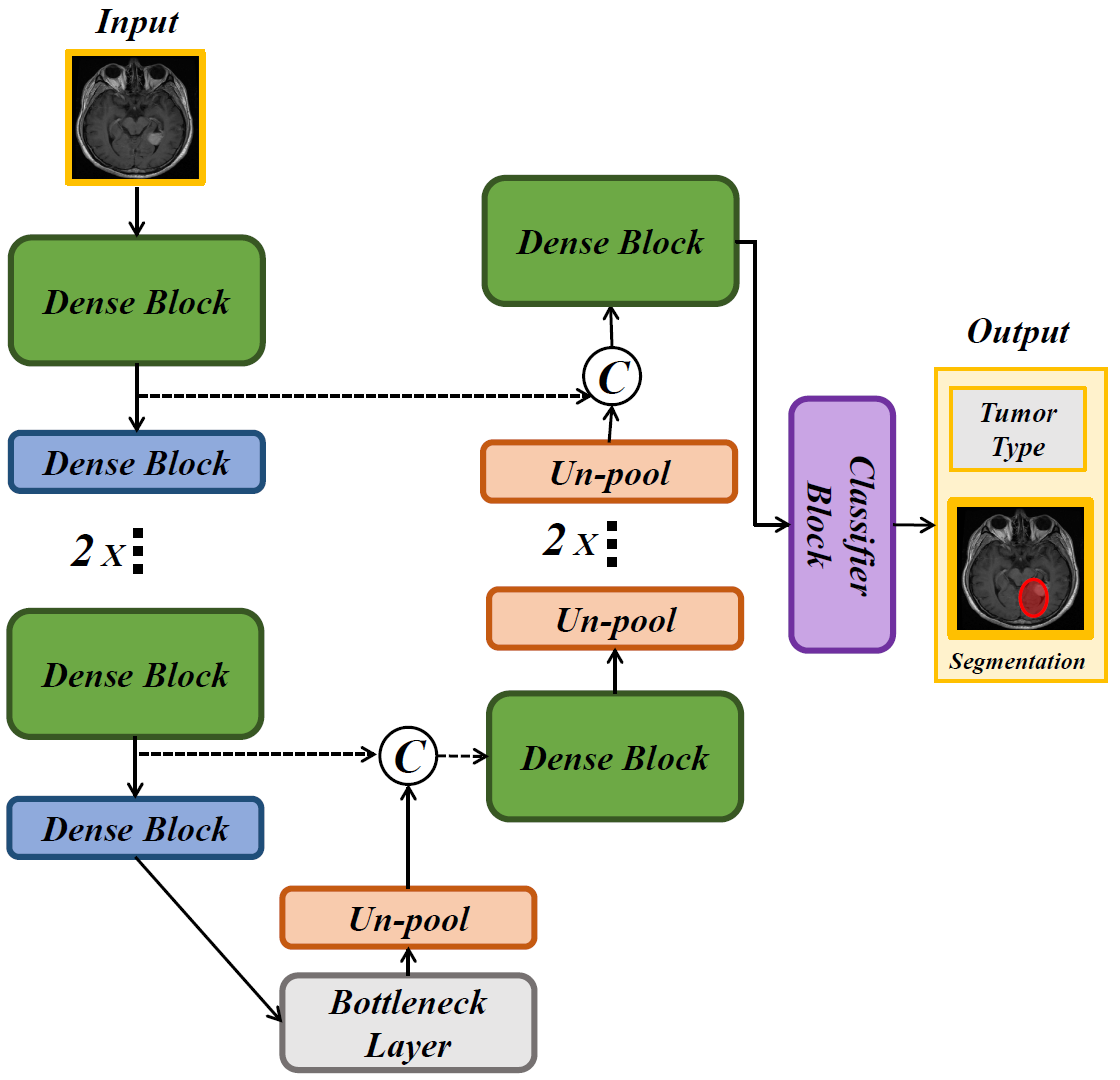}
      \caption{Illustration of network architecture consisting of dense, bottleneck and classifier blocks shown in the left view. the c in circle symbol represent concat in third dimension. Code available at: \href{https://github.com/soroush361/QuickTumorNet.git}{Github Link}  }
      \label{figurelabel}
   \end{figure}
   
\par MRI has enabled the possibility to detect and classify brain tumors non-invasively. However, it is still a laborious task that requires an expert radiologist and is subjective by nature and therefore results in high inter-rater reliabilty variance~\cite{isunuri2020fast}. Additionally, even expert radiologists can only perform a small amount of diagnostic reads, tumor type classifications, and tumor quantifications or segmentations when compared to the amount a CNN can in the same duration. 
\par In the medical discipline of radiology, the use of computer-based image processing techniques for anatomical segmentation and pathological classification have gained popularity as deep learning and machine learning neural networks increase in their accuracy and run-time efficiency~\cite{lundervold2019overview,dong2017automatic}. Tumor segmentation is important as it is in essence detecting the presence of a tumor while quantifying its size. Not only, is segmentation important for tumor detection alone, but for tracking the progress of therapies over time as well. The ability to be able to quantify the size of a brain tumor during treatment helps clinicians know if the administered therapy is effective. Computer aided brain tumor multi-class segmentation is a non trivial task that requires the use of deep neural networks and its classifier’s to learn semantic features of brain tumor types.
\par In this study, we propose a fully automated U-based deep CNN~\cite{noh2015learning}, QuickTumorNet, inspired by QuickNAT's architecture, that demonstrated promising performance of accurate and reliable brain tissue type segmentation with a reasonable run-time.

\begin{figure}[b]
\begin {center}
    \includegraphics[width=0.4\textwidth]{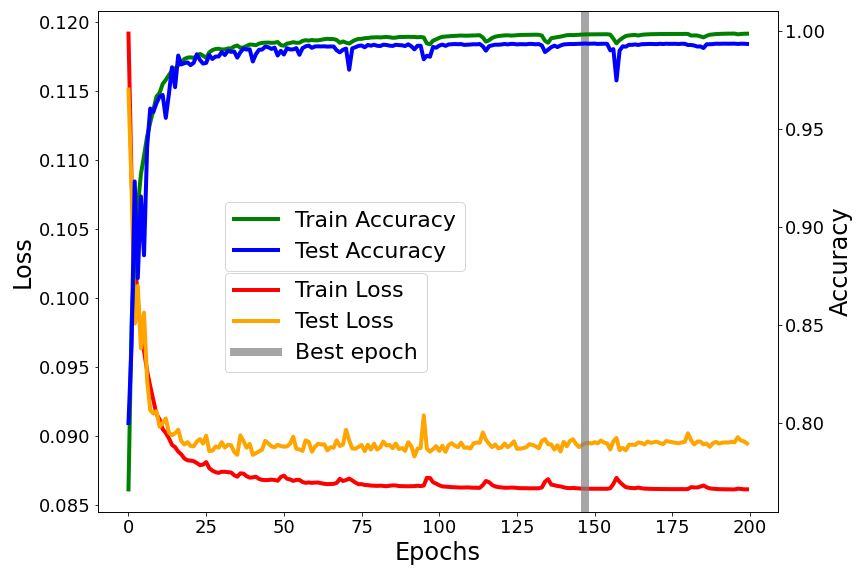}
\caption{QuickTumorNet learning curve}
\label{fig:block_diag}
\end {center}
\end{figure}

\section{Methods}
\subsection{Data Set} 
The brain data set utilized in this study contained 3064 T1-weighted contrast-enhanced (CE) MRI images collected from 233 patients. They were acquired between 2005 and 2010 at The Southern Medical University Nanfang Hospital in Guangzhou, China and The Tianjin Medical University General Hospital in Heping, China. The dataset contained patients diagnosed with three different classes of brain tumors (meningioma, glioma, and pituitary) consisting of 708 slices of meningioma tumors, 1426 slices of glioma tumors, and 930 slices of pituitary tumors. To best simulate the radiologist experience and clinical practice, rather than using a 3D model of the tumor data, we used only a certain number of slices of brain CE-MRI with a large slice thickness~\cite{cheng2015enhanced}. Accordingly, the proposed architecture is based on 2D axial, sagittal, and coronal slices. The slice thickness is 6 mm and the slice gap is 1 mm. Tumor borders were manually delineated and annotated by three experienced radiologists. Slices that contained large tumor sizes were selected to construct the dataset~\cite{cheng2016retrieval,cheng2015enhanced}.

\subsection{Network structure}

In this study, we implemented a modified version of QuickNAT~\cite{roy2019quicknat}, a fully convolutional neural network designed for fast and precise brain tissue type segmentation. The modified network has an encoder/decoder like 2D F-CNN architecture with 4 encoders and 4 decoders separated by a bottleneck layer shown in Fig. 1. The final layer is a classifier block with softmax. The architecture includes skip connections between all encoder and decoder blocks of the same spatial resolution, similar to the U-Net architecture~\cite{ronneberger2015u} which provide encoded feature information to the decoder also providing a path of gradient flow from the shallow layers to the deeper layers, improving training~\cite{he2016deep,drozdzal2016importance}. In the decoding stages, unlike  a classical U-NET that uses up-sampling feature maps for transpose convolution, we employed un-pooling layers~\cite{noh2015learning}.

     \begin{figure*}[t]
      \includegraphics[width=\textwidth]{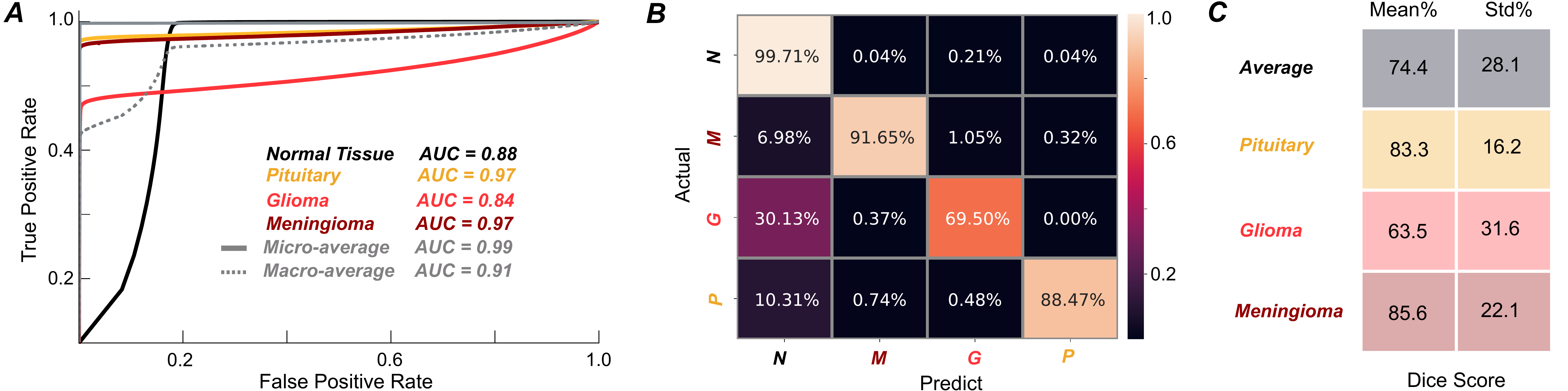}
    \caption{Trained Model Figures and Metrics: Dice score, ROC Curve and, Model Confusion Matrix from left to right respectively.}
\label{figurelabel}
  \end{figure*}

\subsection{Training and Optimization}
The main challenge encountered in training the model was the slow dispersion of error back propagation caused by the low percentile of informative pixels in the data (the number of pixels are two orders of magnitude larger than the number of informative pixels for each class). Two possible ways to address this issue of high precision against low recall involves either preprocessing the images and creating tumor masks to input into the network or apply and take advantage of a more complex loss function \cite{taghanaki2019combo}. To overcome this problem while simultaneously solving the numerical instability issue we used the same algorithm implemented in~\cite{tetteh2018deepvesselnet} where it modified the cross-entropy loss function weighted by additional factors to remedy the high false positive rate as below:

$$
\begin{aligned}
\mathcal{L}(\mathbf{W})=& \mathcal{L}_{1}(\mathbf{W})+\mathcal{L}_{2}(\mathbf{W}) \\
\mathcal{L}_{1}(\mathbf{W})=&-\frac{1}{\left|Y_{+}\right|} \sum_{j \in Y_{+}} \log P\left(y_{j}=1 \mid X ; \mathbf{W}\right) \\
&-\frac{1}{\left|Y_{-}\right|} \sum_{j \in Y_{-}} \log P\left(y_{j}=0 \mid X ; \mathbf{W}\right) \\
\\
\mathcal{L}_{2}(\mathbf{W})=&-\frac{\gamma_{1}}{\left|Y_{+}\right|} \sum_{j \in Y_{f+}} \log P\left(y_{j}=0 \mid X ; \mathbf{W}\right)\\
&-\frac{\gamma_{2}}{\left|Y_{-}\right|} \sum_{j \in Y_{f-}} \log P\left(y_{j}=1 \mid X ; \mathbf{W}\right)
\end{aligned}
$$
$$
\begin{array}{c}
\gamma_{1}=0.5+\frac{1}{\left|Y_{f+}\right|} \sum_{j \in Y_{f+}}\left|P\left(y_{j}=0 \mid X ; \mathbf{W}\right)-0.5\right| \\
\gamma_{2}=0.5+\frac{1}{\left|Y_{f-}\right|} \sum_{j \in Y_{f-}}\left|P\left(y_{j}=1 \mid X ; \mathbf{W}\right)-0.5\right|
\end{array}
$$
\\
$\mathcal{L}$ is the loss function and \textbf{W} represent the network parameters trained by backpropogation. P(.) is the probability of the softmax layer, and $Y+$ and $Y-$ are the set of Tumor and non-Tumor classes.

Training deep neural networks requires stochastic gradient-based optimization to minimize the cost function with respect to its parameters. We adopted the adaptive moment estimator (Adam)~\cite{kingma2014adam}  to estimate the parameters. In general, Adam utilizes the first and second moments of gradients for updating and correcting moving average of the current gradients. The parameters of our Adam optimizer were set as: learning rate = 0.0001 and the maximum number of epochs = 200. All weights were initialized by normal distribution and all biases were initialized as 0.

\section{Results}

\subsection{Experiments and Performance Evaluation}
The model presented, QuickTumorNet performed exceedingly well in brain tumor tissue segmentation as well as multi-classification of the presented brain tumor classes demonstrated by the following metrics and figures [Figure 2 - QuickTumorNet's Learning Curve, Figure 3a-c: ROC Curve - Figure 3a, Confusion Matrix - Figure 3b, and DICE scores - Figure 3c].

\subsubsection{Learning Curve} 
The learning curve of QuickTumorNet, produced in Figure 2, indicates the stability of the network as it's accuracy/learning converges with no significant loss or fluctuation in it's training accuracy past epoch 75. In addition to the network's stability, we can extrapolate that the train accuracy reliably predicts it's test accuracy. The maximum accuracy achieved on the test data set was an excellent percentage of 99.35\%. 

\subsubsection{The Area Under the Curve (AUC)}
The AUC of a ROC plot is an estimate of the probability that a classifier will rank a randomly chosen positive instance higher than a randomly chosen negative instance. Referring to Figure 3a, the AUC for non-tumorous brain (normal) tissue is 0.88, for meningioma tissue is 0.97, for glioma tissue is 0.84, and for pituitary tissue is 0.97 indicating that a classifier in the model has a very high chance of ranking a random positive instance over a random negative instance for each class. The AUC for pituitary tissue and meningioma tissue are both an outstanding 0.97. For normal tissue and glioma tissue, their AUC are both in the 0.8 range, 0.88, and 0.84 respectively, indicating the model built and its classifiers are performing very well.       

\subsubsection{Confusion Matrix}
The Confusion Matrix, Figure 3b is informative of the model’s accuracy in both segmentation and multi-classification between the three label classes as well as the percent of one class mislabeled as another. QuickTumorNet correctly labeled normal tissue with an accuracy of 99.71\%. This is a remarkable percentage as it means in the model’s overall segmentation task it correctly segmented or identified 99.71\% of the normal tissue and only incorrectly predicted and mislabeled 0.28\% of normal tissue as tumorous tissue. 
\par The model predicted and labeled meningioma tissue with an accuracy of 91.65\% and pituitary tissue with an accuracy of 88.47\%. Accuracy of 99.71\% and 91.65\% for meningioma and pituitary tissue respectively are exceptional results for a multi-class segmentation task. While the lowest accuracy for an individual tumor class for glioma tissue of 69.50\% is still an exciting result, especially in addition to the superb accuracy seen for meningioma and pituitary tumors, it is worth noting that this may be be due to the nature of glioma tumor's themselves and/or the specific data set provided to the network. Both of which will be further discussed later on.

\subsubsection{DICE Scores}
We calculated the mean and standard deviation of the DICE scores for each tumor type individually and for all of the tumor types averaged together as seen in Figure 3c. DICE scores measure the extent of the overlap of the true label annotations and the model’s predicted segmentation. In this instance, in multi-class segmentation, the DICE scores measure the extent of the overlap of the three annotated classes and the model’s predictions. DICE scores, not only provide further insight into the performance of our model but can also offer some explanations into its shortcomings which can help guide future work and improvement. 
\par We calculated the mean and standard deviation of the DICE scores for each tumor class to demonstrate the possible explanation for the model’s performance on glioma tissue on average was around 20 percent lower than the other two tumor classes with a much larger standard deviation. Again, this further supports that the model’s performance and accuracy on glioma tissue may be due to the nature of glioma tumors themselves or the specific data set employed.

   
     \begin{figure*}[t]
      \includegraphics[width=\textwidth]{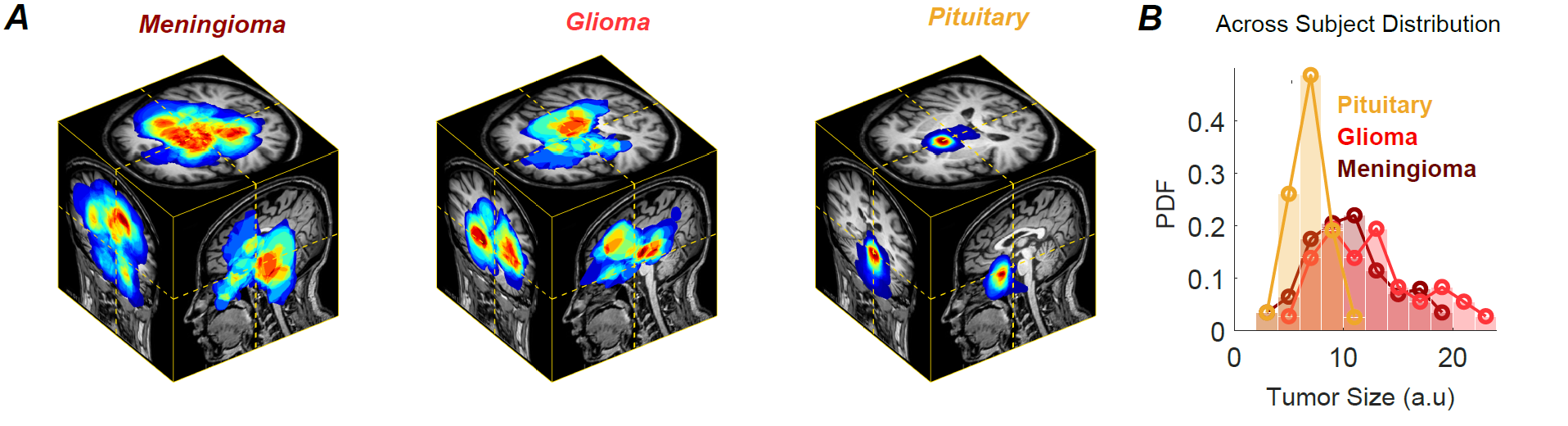}
    \caption{Spatial Distribution of tumors across patients presented in blue-red false color. Tumor class size distributions across data set.}
\label{figurelabel}
  \end{figure*}
\section{Discussion}

\subsection{Spatial Distribution and Sizes of Tumor classes in Data Set}
Produced in Figure 4a is the spatial distributions of each tumor class provided to QuickTumorNat rendered as a 3D image via manual traditional registration- the process of transforming different sets of data into one coordinate system ~\cite{hariri2015medical}- with 9 degrees of freedom. Figure 4b is the spatial distribution of the sizes of each tumor class provided to our model.
\par Figure 4a left and middle show the wide range and variability in brain tumor sizes and locations of tumor tissues belonging to meningioma and glioma brain tumor classes. Figure 4b, the similarity of the distribution in sizes across the data set of meningioma and glioma tumor classes is seen. This collectively indicates the difficulty any deep learning neural network would have distinguishing between meningioma and glioma tumor classes based on size and location alone in this data set. 
\par Despite this, our network was able to accurately predict and correctly label meningioma tumor tissue with an accuracy of 91.65\% while still classifying glioma tumor tissue with an accuracy of almost 70\%. This indicates our model is learning semantic features about the tumor texture patterns of the different tumor tissue classes. 
\subsection{Visual Examples of Highest and Lowest DICE SCORES}

\subsubsection{Highest DICE Scores}Included in Figure 5 are the three highest DICE scores from each tumor class from the three planes totaling nine images/results. From top to bottom, row one are tumors belonging to the meningioma class, row two are tumors belonging to the glioma class, and row three are tumors belonging to pituitary class. The green boundaries are the expert radiologist's annotations used as ground truth segmentation's for each class. Blue boundaries are QuickTumorNet's predicted labels for each tumor class when it correctly predicted and classified the respective tumor tissue class. Red boundaries were used when the model predicted and segmented tumor tissue but miss-classified them as belonging to another class.

\par It is clear how well the model performed in these nine instances. Firstly, the lack of presence of any red boundaries tells us there are no miss-classifications between tumor tissue types. The blue and green boundaries virtually line up in agreement demonstrating not only the network's performance in this specific data set but its potential to perform on more complete varied data sets.

\subsubsection{Examples of Lowest DICE Scores}
Organized in the same fashion as Figure 5, Figure 6 are the three lowest DICE scores from each tumor class from each plane. It is to be expected that the three lowest DICE scores from each class and plane, that the model would not predict and overlap nearly as well with the annotations as it did in the previous figure of the highest DICE scores, Figure 5. It is worthwhile to note however that the model did still predict and correctly label the appropriate tumor class in a vast majority of the lowest DICE scores (eight out of nine results). Even on the result where it did not correctly label the appropriate class, the coronal meningioma slice, it still did not lose the presence of tumor tissue completely as indicated by the presence and overlap of the red boundary with the green boundary.

\subsection{Varied Nature of Tumor's and Data Set Utilized}
The nature and diversity of each tumor class, particularly glioma tumors, as their occurrences are so diverse and vast ~\cite{liang2020prognostic} is one particular possible explanation for the model's relatively less satisfying performance on glioma tissue. Additionally, there are rare cases and presentations of all tumor classes that may have been included in the data set that negatively impacted its overall performance. There is also always the possibility that the radiologist's miss-classified the tumor type. 
 \begin{figure}[h]
      \centering
      \includegraphics[width=0.4\textwidth]{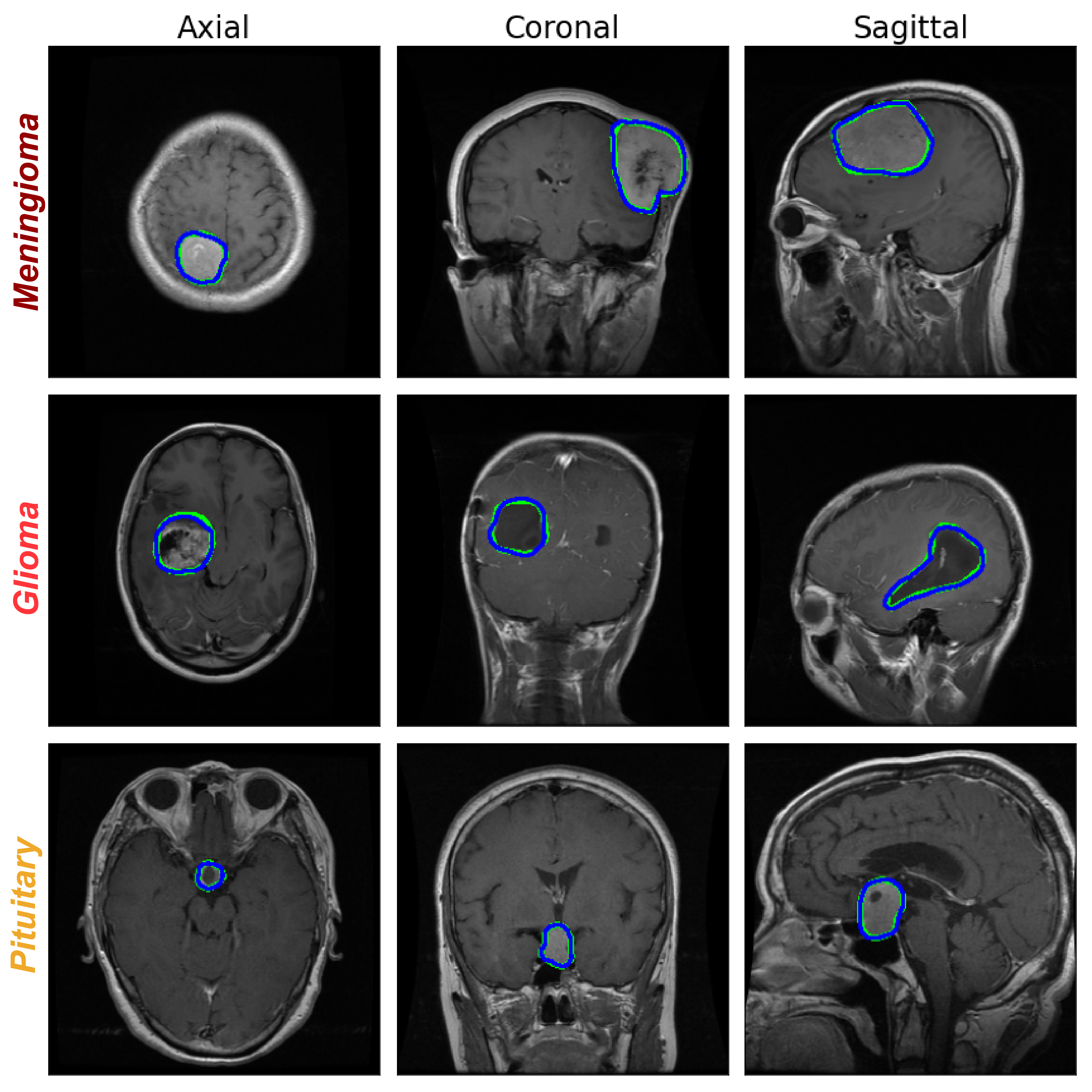}
      \caption{Network Segmentation - High DICE Scores: Three brain tumor classes in the three different planes from our local test set. Green delineations are the radiologist annotations (ground truth) and blue borders are the network's predicted multi-class segmentations correctly classified.}
      \label{figurelabel}
   \end{figure}

\par Although the number of samples for glioma tumors in the data set is the largest, the nature of the progression of glioma tumors and their presentation on MRI acquired scans changes throughout the stages of glioma tumor progression. While the other two tumor tissue classes, meningioma and pituitary, remain relatively consistent in their presentation on MRI scans, glioma tumors intensities change throughout the progression and stages of the tumor~\cite{liang2020prognostic, sarbu2016increased}. This offers a possible explanation as to why we are seeing high mean DICE scores for meningioma ($85.6\pm 22.1\%$) and pituitary tumors ($83.3 \pm 16.2\%$) compared to glioma tumors ($63.5 \pm 31.6$). Furthermore, given the low standard deviation associated with meningioma and pituitary tumor classes, we may be able to infer that the high standard deviation seen with glioma tumors is caused by the size and location variation and their presentation at different stages of disease progression in our selected data set.

         \begin{figure}[t]
      \centering
      \includegraphics[width=0.4\textwidth]{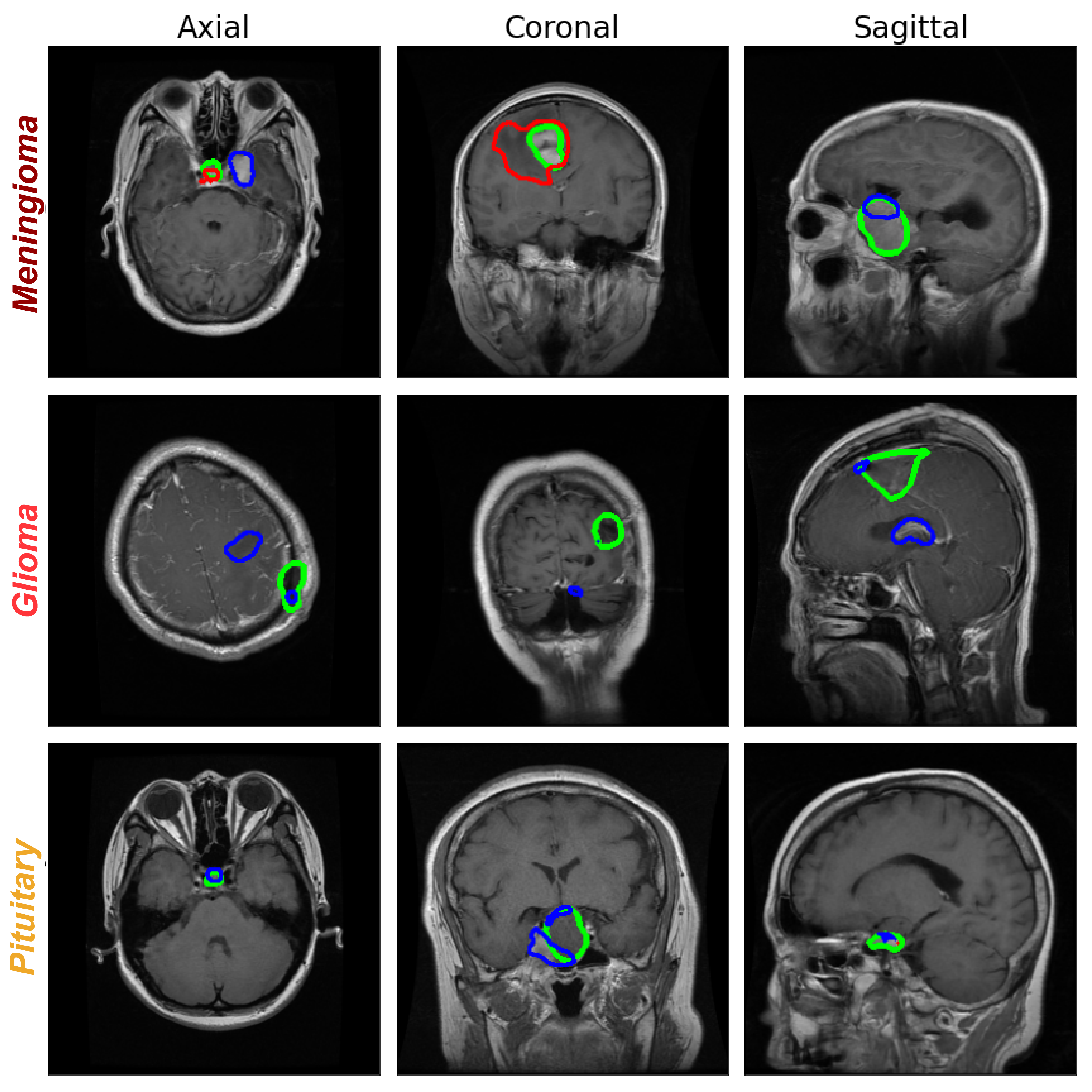}
      \caption{Network Segmentation - Low DICE Scores: (Color coding similar as Figure 5)}
      \label{figurelabel}
   \end{figure}
\par 
\section{Conclusion}
Currently, while no CNN model is perfect, QuickTumorNet demonstrated promising results yielding reliable and accurate brain tumor segmentation and multi-class segmentation with fast performance and processing speed of 19 milliseconds per slice. Due to the subjectivity involved in manual radiological reads, false classification and quantification can occur where it is advantageous to have a "second opinion" on difficult cases from a segmentation CNN~\cite{cheng2015enhanced} such as QuickTumorNet.
\section{Possible Future Work}
Possible future work includes image preprocessing such as brain extraction, image registration, voxel intensity thresholding or voxel intensity scaling. Training separate networks for each plane individually (i.e. the axial plane, the coronal plane, and the sagittal plane).
Comparing the network model to simpler architectures can be beneficial to understand underlying information that network is extracting from medical images~\cite{zhuang2020classification}.
The use of additional data sets covering more varied samples and instances of rare occurrences as well more diversified cohorts.

\addtolength{\textheight}{-12cm}   



\bibliographystyle{unsrt}  
\bibliography{root}

\end{document}